\documentclass[pdflatex,sn-mathphys,Numbered]{sn-jnl}

\usepackage{graphicx}%
\usepackage{multirow}%
\usepackage{amsmath,amssymb,amsfonts}%
\usepackage{amsthm}%
\usepackage{mathrsfs}%
\usepackage[title]{appendix}%
\usepackage{xcolor}%
\usepackage{textcomp}%
\usepackage{manyfoot}%
\usepackage{booktabs}%
\usepackage{algorithm}%
\usepackage{algorithmicx}%
\usepackage{algpseudocode}%
\usepackage{listings}%

\usepackage[caption=false,font=normalsize,labelfont=sf,textfont=sf]{subfig}
\usepackage[commandnameprefix=ifneeded,final]{changes} 
\definechangesauthor[name={S.Li}, color=red]{A}

\begin{document}

\title[Seismogram Transformer]{SeisT: A foundational deep learning model for earthquake monitoring tasks}


\author[1]{\fnm{Sen} \sur{Li}}\email{sli@cumt.edu.cn}

\author*[1]{\fnm{Xu} \sur{Yang}}\email{yang\_xu@cumt.edu.cn}

\author*[2]{\fnm{Anye} \sur{Cao}}\email{caoanye@163.com}

\author[3]{\fnm{Changbin} \sur{Wang}}\email{changbin.wang@cumt.edu.cn}

\author[2]{\fnm{Yaoqi} \sur{Liu}}\email{yaoqi\_liu@cumt.edu.cn}

\author[1]{\fnm{Yapeng} \sur{Liu}}\email{liuyp@cumt.edu.cn}

\author[1]{\fnm{Qiang} \sur{Niu}}\email{niuq@cumt.edu.cn}

\affil*[1]{\orgdiv{School of Computer Science and Technology}, \orgname{China University of Mining and Technology}, \orgaddress{\street{Daxue Road}, \city{Xuzhou}, \postcode{221116}, \state{Jiangsu}, \country{China}}}

\affil[2]{\orgdiv{School of Mines}, \orgname{China University of Mining and Technology}, \orgaddress{\street{Daxue Road}, \city{Xuzhou}, \postcode{221116}, \state{Jiangsu}, \country{China}}}

\affil[3]{\orgdiv{State Key Laboratory of Coal Resources and Safe Mining}, \orgname{China University of Mining and Technology}, \orgaddress{\street{Daxue Road}, \city{Xuzhou}, \postcode{221116}, \state{Jiangsu}, \country{China}}}

\abstract{Seismograms, the fundamental seismic records, have revolutionized earthquake research and monitoring. Recent advancements in deep learning have further enhanced seismic signal processing, leading to even more precise and effective earthquake monitoring capabilities. This paper introduces a foundational deep learning model, the Seismogram Transformer (SeisT), designed for a variety of earthquake monitoring tasks. SeisT combines multiple modules tailored to different tasks and exhibits impressive out-of-distribution generalization performance, outperforming or matching state-of-the-art models in tasks like earthquake detection, seismic phase picking, first-motion polarity classification, magnitude estimation, back-azimuth estimation, \added{and epicentral distance estimation}. The performance scores on the tasks are 0.96, 0.96, 0.68, 0.95, 0.86, 0.55, \added{and 0.81}, respectively. The most significant improvements, in comparison to existing models, are observed in phase-P picking, phase-S picking, and magnitude estimation, with gains of 1.7\%, 9.5\%, and \added{8.0\%}, respectively. \added{Our study, through rigorous experiments and evaluations, suggests that SeisT has the potential to contribute to the advancement of seismic signal processing and earthquake research.}}




\maketitle

\section{Introduction}\label{sec1}
Deep learning approaches have gained traction in various seismological subfields, exhibiting promising capabilities that often surpass traditional methods and lead to notable performance enhancements \cite{RN18_0912_1, RN172_0912_2, RN13_0912_3,mohammadigheymasi2023ipiml,wang2022unsupervised}. Their efficiency and robust performance have made deep learning-based tools in seismology increasingly popular in various research domains.

\added{Recorded seismic waveforms serve as the starting point for seismicity monitoring and form the basis for a series of subsequent analyses. Deep learning models acquire compact representations of seismic signals from large-volume seismic datasets}, enabling them to extract deep-level feature representations of seismic records. \added{Notably, in the domain of seismology, significant advancements have been achieved through deep learning methodologies applied to single-station scenarios \cite{Mousavi_MachineLearningInEarthquakeSeismology}. These advancements encompass diverse aspects, including} earthquake detection \cite{RN90_0912_4, RN11_0912_5, RN255_0912_6,liao2022red}, phase picking \cite{RN11_0912_5, RN255_0912_6, RN266_0912_7, RN257_0912_8, liao2022red}, first-motion polarity classification \cite{RN311_0912_9, RN85_0912_10, RN115_0912_11}, magnitude estimation \cite{RN17_0912_12, RN357_0912_13}, back-azimuth estimation \cite{RN51_0912_14, RN359_0912_15}, and \added{epicentral distance estimation \cite{RN51_0912_14, Ristea2022ComplexNN}. However, research on foundational models in seismology is still in its early stages, with only a few relevant models recently proposed, such as SeisCLIP \cite{si2023seisclip} and SFM \cite{sheng2023SFM}. A foundational model \cite{bommasani2022opportunities}, trained on large-volume seismic datasets and capable of handling various downstream tasks related to seismogram analyzing, remains scarce. Currently, task-specific models are the norm for automation of seismogram processing. However, this could result in a lack of uniformity and efficiency in trained models.} Meanwhile, in artificial intelligence fields such as Computer Vision (CV) and Natural Language Processing (NLP), various robust foundational network models have emerged, achieving significant success in their respective domains. Hence, this study posits the existence of a feature space suitable for multiple related seismological tasks, effectively representing seismic waveform information. This transformation allows the conversion of various seismic waveform analysis tasks into the problem of fine-tuning this feature space for specific tasks.

\added{In this paper, a foundational model is designed, which functions as a feature extractor. Convolutional neural networks possess translational invariance and local induction bias, rendering them highly effective for extracting local features.} However, they may lack the ability to establish long-term dependencies \cite{RN290_0912_17}, which may be necessary for certain seismic waveform analysis tasks. To overcome this limitation, we introduce the global self-attention mechanism of the Transformer to capture interactions between global and local features \cite{RN361_0912_18}. \added{Therefore, the design of this foundational model represents a hybrid network architecture combining Transformer \cite{RN270_0912_16or32} and convolution.}

Furthermore, we observed challenges in the out-of-distribution generalization capabilities of existing models across different regions and seismic networks. Challenges may arise due to the differences in geology, varying epicentral distances, and other factors leading to differences in seismic waveform features. \added{Inspired by the Short-Term Average to Long-Term Average (STA/LTA) algorithm \cite{RN363_0912_19}, we designed Multi-Scale Mixed Convolution (MSMC) modules, Local Aggregation (LA) modules, and Multi-Path Transformer (MPT) modules to enhance the generalization capabilities of the model on different data distributions.} Additionally, various output head modules are also designed, making this foundational network suitable for various tasks, including but not limited to earthquake detection, phase picking, first-motion polarity classification, magnitude estimation, back-azimuth estimation, \added{and epicentral distance estimation}.

In this paper, a novel neural network architecture called the Seismogram Transformer (SeisT) is proposed to address a key challenge in the application of deep learning to \added{earthquake monitoring}: the construction of a foundational network to enhance the efficiency and consistency of seismic waveform analysis tasks while ensuring generalization capabilities. We conducted model training on seismic data from mainland China and evaluated the performance of the model on seismic records from both mainland China and the Northwestern Pacific region of the United States, comparing it with state-of-the-art models for various tasks. \added{The results demonstrate that SeisT consistently matches or exceeds the performance of existing approaches, particularly in terms of out-of-distribution generalization, for earthquake detection, phase picking, first-motion polarity classification, magnitude estimation, back-azimuth estimation, and epicentral distance estimation tasks.} This model is of great reference to making significant advancements in seismological research and providing strong support for practical applications such as earthquake early warning and seismic monitoring.

\section{Methods}\label{sec2}

\subsection{Network architecture}\label{subsec2.1}

\begin{figure*}[ht]

\centering
\includegraphics[width=16cm]{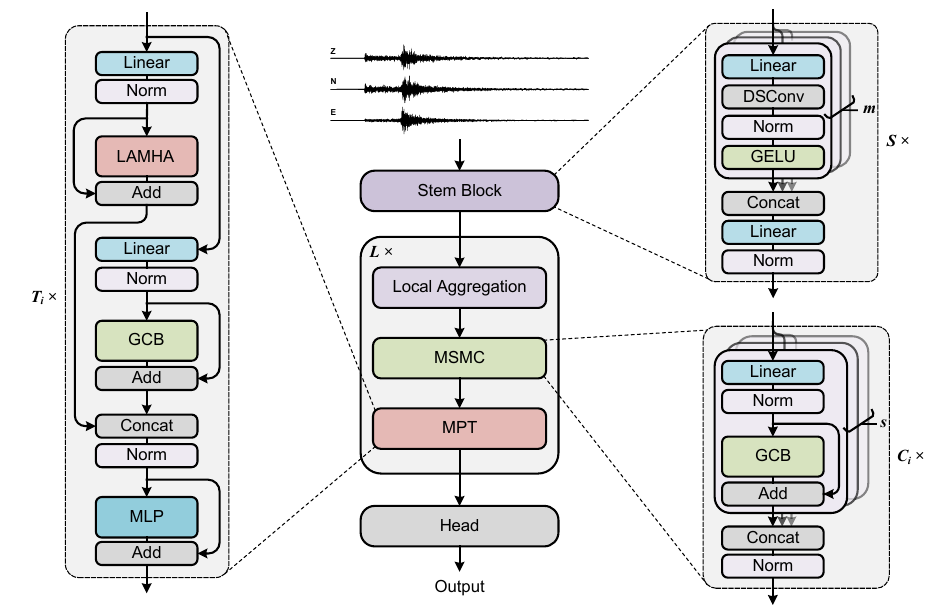}
\caption{SeisT model architecture. \added{The proposed model takes the normalized 3-component seismic waveform as input}, preprocesses it through a stem module containing $S$ stem layers, and then maps it to a high-dimensional feature space through $L$ body modules. The $i$-th body module consists of a local aggregation module, $C_i$ MSMC modules, and $T_i$ MPT modules. \added{It incorporates several fundamental modules, such as Group Convolution Block (GCB), Local Aggregation Multi-Head Attention (LAMHA), Depthwise Separable Convolution (DSConv), and Multi-Layer Perceptron (MLP).} The model outputs vector representations specific to the given task through the head module. } 
\label{fig: fig1}
\end{figure*}

The proposed novel network architecture (Fig. \ref{fig: fig1}) consists of three main components: stem, body, and head. \added{The stem performs initial processing on the input normalized seismograms by utilizing multi-path depthwise separable convolution and linear transformations. Through multiple layers of processing, it swiftly reduces the dimensionality of the input seismograms from $l$ to $l/4$ along the time axis, aiming to alleviate the computational load of subsequent attention modules while obtaining fundamental feature representations. The body section comprises $L$ stages, with each stage containing three parts: LA module, MSMC modules, and MPT modules.} This structure enables the effective extraction of deep-level features. \added{The head module is responsible for decoding the extracted features and outputting the target predictions.} Multiple heads tailored for different tasks are designed, allowing the stem and body to share the same feature map across various tasks. The flexible modular design enables the model to be adapted to different types of tasks, including earthquake detection, seismic phase picking, first-motion polarity classification, magnitude estimation, back-azimuth estimation, \added{and epicentral distance estimation}. \added{Due to the combined structure of convolution and self-attention, the network architecture achieves independence from the length of seismograms during its construction. SeisT can handle seismograms of varying lengths without the need for predefined lengths during training and inference. Specifically, in the convolutional layers, it can automatically pad to match the convolution kernel size, and during up-sampling, it can dynamically calculate the up-sampling ratio for each stage based on the dimensions of the input seismograms. For regression and classification tasks, global average pooling is utilized to integrate feature maps of different shapes.}

\added{The stem and body modules collaboratively work to map the seismic waveforms into a feature space with compressed time dimensions and increased channel depth. This contributes to capturing a broader range of temporal information and the complex structures and patterns within the input data}. Within the body, the multi-head self-attention structure plays a crucial role, which efficiently captures long-range dependencies, thus mitigating the limitations of receptive fields in convolutional networks. In the self-attention mechanism, for seismograms with a length of $L$ and a channel size of $C$, the algorithm complexity is exceedingly high, reaching $\mathcal{O}(L^{2}C)$. \added{Therefore, the combination of local aggregation and self-attention mechanisms reduces the dimensions of key and value, significantly reducing computational overhead without decreasing the output dimensionality. At the beginning of each stage, there is a local aggregation module responsible for down-sampling the temporal dimension of the feature maps, performing local fusion on the feature map with a specified window size for each stage.} These down-sampled features undergo feature extraction through a series of MSMC modules, resulting in more advanced and rich feature representations. Subsequently, these feature maps are inputted into one or more concatenated MPT modules, allowing deep-level features to be fused globally through attention mechanisms. Each module includes multiple residual connections and normalization layers, which facilitate the convergence of deep neural networks and reduce training difficulties.

For various tasks, three types of head structures have been designed. \added{The first type of head is designed for earthquake detection and phase picking, composed of multiple consecutive up-sampling stages.} Each stage includes linear interpolation and convolution operations to restore multiple down-sampled feature map sizes corresponding to the stem and body. Then, the convolution modules and the sigmoid function are employed to integrate the output, mapping high-level feature information into three probability sequences related to the existence of seismic signals at each time point, phase-P, and phase-S. \added{The second type of head is designed for classification tasks utilized for first-motion polarity classification. It incorporates global average pooling to aggregate features and reduces computational complexity, enabling the head to adapt to input feature maps with different sizes in the time dimension. Subsequently, a linear transformation and the softmax function are applied to output a vector of dimension 2.} \added{The third type of head is designed for regression tasks. Similar to the aforementioned second type (classification head), it shares commonalities but differs in that the final output undergoes sigmoid scaling instead of utilizing softmax normalization. This type of head is employed for regression tasks such as back-azimuth estimation, magnitude estimation, and epicentral distance estimation.}

\added{The SeisT designed in this paper is classified into three sizes based on its number of layers: Small (S), Medium (M), and Large (L). This categorization is intended to meet the requirements of different application scenarios and varying training data volumes. Including the head, the smallest network has a total parameter count of only 98k, while the largest network does not exceed 670k.} The architecture of these networks draws inspiration from several visual and natural language processing models \cite{RN177_0912_26, RN293_0912_27, RN316_0912_28, RN371_0912_29, RN373_0912_30} and is designed based on domain-specific knowledge \added{in seismology}. The selection of model structure and hyperparameters is informed by extensive experimentation with numerous prototype networks (Supporting Table S1).

\subsection{Network design}\label{subsec2.2}
During seismic events, seismic signals \added{are typically captured} by multiple stations at varying distances and of different types. Due to differences in propagation paths and travel times, the signal characteristics received by various stations in the seismic network exhibit certain disparities. To address the variations in the temporal dimension, we introduced grouped convolutions with multiple different-sized kernels to allow the network to flexibly extract local signal features under various receptive field conditions. This helps mitigate the information loss caused by a single receptive field, allowing the model to handle the sample differences present in the temporal dimension. Convolutional neural networks possess a strong inherent bias for efficiently extracting local signal features. However, for tasks such as earthquake detection that require a focus on global features, convolutional networks may constrain the interrelation of global features. Transformer modules have been proven to be an effective architecture for seismic signal processing and modeling \cite{RN11_0912_5}. Its self-attention mechanism promotes the fusion of global features. \added{In the original Transformer architecture, each token represents a single word. When applied to the seismograms, the information content of an individual sampling point is minimal. In processing seismic waveforms, one token corresponds to multiple samples at a given moment in time. Therefore, we designed the local aggregation module to represent local features by aggregating information from multiple nearby samples.} This allows for interactions between local features, replacing point-to-point feature interactions. When the LA module is applied to the Transformer layer, it can reduce the dimensionality of key and value in the multi-head self-attention mechanism, thus reducing time complexity.

To further reduce information loss and computational overhead, the MPT module was designed to parallelize the Transformer and convolution operations effectively. This module reduces the computational load of multi-head self-attention and supplements global fusion features with the local feature extraction capabilities of convolutional neural networks. The foundational model employs the stem block for preprocessing seismic signals to adapt them for subsequent deep feature extraction. Various types of output heads were designed for different tasks (Supporting Figure S1), enabling the flexible application of this foundational model to multiple tasks related to seismic signals.

\subsection{Multi-scaled mixed convolution}\label{subsec2.3}
In order to more efficiently model \added{seismic waveforms} and features with the existing convolutional modules, we designed an efficient and adaptable convolutional module called the Multi-Scale Mixed Convolution. This module enhances the feature generalization ability of representation and accelerates inference speed. With inspiration from various modules such as Inception \cite{RN371_0912_29} and Xception \cite{RN375_0912_31}, we project the input shallow-level features through linear projection into multiple feature spaces, denoted as $\{\mathcal{X}_{i}\in \mathbb{R}^{d_i\times t} \mid i=1,2,\cdots,s\}$, where $d_i$ represents the number of output channels of the $i$-th sub-convolution module; \added{$t$ represents the length of the time dimension; $s$ represents the total number of sub-convolution modules.} \added{This approach enables the model to consider feature space information from diverse scales and positions, promoting the effective learning of representations.} The proposed MSMC is defined as follows:

\begin{equation}
    \text{MSMC}(\mathbf{X})=\text{BN}(\mathcal{C}(\mathbf{S}_1,\mathbf{S}_2,\cdots,\mathbf{S}_s))\text{,}
    \label{eq: eq1}
\end{equation}
where,
\begin{equation}
    \mathbf{S}_i=\text{GCB}(\text{BN}(\mathbf{X}^{\top} \mathbf{W}_i+\mathbf{b}_i))\text{.}
    \label{eq: eq2}
\end{equation}

Here, $\mathcal{C}(\cdot)$ is the concatenation operation; $X$ represents the input to the MSMC. We perform concatenation along the channel axis to merge all features and normalize them through BatchNorm, denoted as $\text{BN}(\mathbf{X})=\frac{\mathbf{X}-\mathbb{E}(\mathbf{X})}{\sqrt{\text{Var}(\mathbf{X})+\epsilon}}\cdot\gamma+\beta$. $\gamma$ and $\beta$ are learnable affine transformation parameters, where $\epsilon$ is a small constant. $\mathbf{S}_i$ represents the output of the $i$-th sub-convolution module; $\mathbf{W}_h \in \mathbb{R}^{c \times d_h}$ is the learnable parameter matrix for the $i$-th linear transformation; and $\mathbf{b}_i \in \mathbb{R}^{d_i}$ is the bias. $\text{GCB}(\cdot)$ denotes the operation of the proposed grouped convolution block (Supporting Figure S2), which is computed as follows:
\begin{equation}
    \text{GCB}(\mathbf{X})=\mathbf{X}_r+\text{GELU}(\text{BN}(\mathbf{X}_r)^{\top}\mathbf{W}_1+\mathbf{b}_1)^{\top}\mathbf{W}_2+\mathbf{b}_2\text{,}
    \label{eq: eq3}
\end{equation}
where,
\begin{equation}
\mathbf{X}_r=\mathbf{X}_o+\mathbf{X}\text{,}\label{eq: eq4a}
\end{equation}
\begin{equation}
\mathbf{X}_o=\text{GELU}(\text{BN}(\Psi_g(\mathbf{X})))^{\top}\mathbf{W}_o+\mathbf{b}_o\text{.}\label{eq: eq4b}
\end{equation}

Here, $\text{GELU}(x)=0.5x(1+\text{tanh}(\sqrt{2/\pi}(x+0.044715x^3))$; $\Psi_g(\cdot)$ represents the standard grouped convolution operation. Experimental results from ablation studies indicate that the MSMC outperforms standard convolutional operations.

\subsection{Multi-path Transformer}\label{subsec2.4}
Transformer has demonstrated outstanding performance in addressing the problem of long-range dependencies \cite{RN270_0912_16or32}. However, its computational complexity reaches $\mathcal{O}(N^2)$, which is not feasible for long sequences of seismic wave inputs. \added{Considering the efficiency and generalization capabilities of large neural networks, we designed the MPT module to incorporate local aggregation and convolution while maintaining a sparse structure.} This sparse architecture helps reduce computational demands and mitigate overfitting issues. The similarity matrices in the Transformer have been \added{proven} to be low-rank matrices \cite{RN377_0912_33}, suggesting that information from similarity matrices can be obtained by a small number of largest singular values. Hence, key and value can be mapped to lower-dimensional subspaces to reduce computational costs.

\begin{figure}[h]
\centering
\includegraphics[scale=0.85]{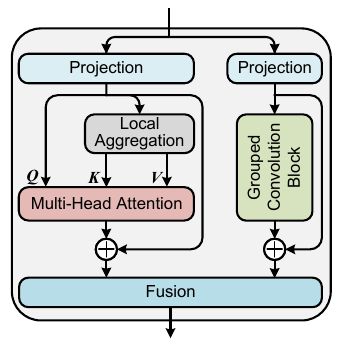}
\caption{Multi-path Transformer architecture. First, the input tensor is projected into two distinct lower-dimensional subspaces. Subsequently, it is separately fed into the Transformer with local aggregation and the grouped convolution block parallel architecture. Finally, feature fusion is achieved through concatenation with a multi-layer perceptron.}
\label{fig: fig2}
\end{figure}

The MPT module employs a parallel structure design (Fig. \ref{fig: fig2}), where the input tensor is linearly projected into two distinct lower-dimensional subspaces. These subspaces then enter a self-attention module with local perception and a grouped convolution module separately. Some prior studies utilized operations such as convolution and pooling to further reduce computational complexity \cite{RN379_0912_34, RN286_0912_35, RN190_0912_36}. In our MPT module, we utilized the proposed local aggregation module, which aggregated the key and value of the multi-head self-attention through local aggregation, enabling each token to incorporate contextual relationships in the temporal dimension.

Given the kernel size $k_a$ for local aggregation, for a tensor with shape $c \times l$, the transformation through local aggregation results in a shape $c \times \lceil l/k_a \rceil$, thereby reducing the computational cost of the similarity matrix. It is then restored to its original shape when combined with a query through dot-product attention. The computation process for multi-head self-attention with local aggregation can be represented as:

\begin{equation}
    \text{LAMHA}(\mathbf{X})=\mathcal{C}(\mathbf{H}_1,\mathbf{H}_2,\cdots,\mathbf{H}_h)^{\top}\mathbf{W}_{ao}+\mathbf{b}_{ao}\text{,}
    \label{eq: eq5}
\end{equation}
where,
\begin{equation}
    \mathbf{H}_i=\text{softmax}(\frac{{\mathbf{Q}_i^{\top}}\mathbf{K}_i}{\sqrt{\mathbf{d}_h}})\mathbf{V}_i\text{,}
    \label{eq: eq6a}
\end{equation}
\begin{equation}
    \mathbf{Q}_i=\mathbf{X}^{\top}\mathbf{W}_{i}^{Q}+\mathbf{b}_{i}^{Q}\text{,}
    \label{eq: eq6b}
\end{equation}
\begin{equation}
    \mathbf{K}_i=\text{BN}(\text{LA}(\mathbf{X},k_a))^{\top}\mathbf{W}_{i}^{K}+\mathbf{b}_{i}^{K}\text{,}
    \label{eq: eq6c}
\end{equation}
\begin{equation}
    \mathbf{V}_i=\text{BN}(\text{LA}(\mathbf{X},k_a))^{\top}\mathbf{W}_{i}^{V}+\mathbf{b}_{i}^{V}\text{,}
    \label{eq: eq6d}
\end{equation}

\added{where $\mathcal{C}(\cdot)$ is the concatenation operation;} $\mathbf{H}_i$ represents the output of the $i$-th head; $\mathbf{W}_{ao} \in \mathbb{R}^{d_a \times d_a}$ and $\mathbf{b}_{ao} \in \mathbb{R}^{d_a}$ are the learnable parameter matrices and bias parameter vectors for the linear projection layer before the output; $\mathbf{Q}_i$, $\mathbf{K}_i$ and $\mathbf{V}_i$ respectively represent the query, key and value vectors of the $i$-th head; $\mathbf{W}_{i}^{Q} \in \mathbb{R}^{d_a \times d_h}$, $\mathbf{W}_{i}^{K} \in \mathbb{R}^{d_a \times d_h}$ and $\mathbf{W}_{i}^{V} \in \mathbb{R}^{d_a \times d_h}$ are the learnable parameter matrices of the $i$-th head respectively; $\mathbf{b}_{i}^{Q} \in \mathbb{R}^{d_h}$, $\mathbf{b}_{i}^{K} \in \mathbb{R}^{d_h}$ and $\mathbf{b}_{i}^{V} \in \mathbb{R}^{d_h}$ are the learnable bias parameter vectors of the $i$-th head, respectively; and \added{$\text{LA}(\cdot,\cdot)$ represents the local aggregation operation.}

The grouped convolution block includes multiple modules such as the convolution layer, linear projection layer, and MLP, which are used to assist Transformer feature extraction. \added{The last two parts of features are spliced and an MLP is used to fuse the features}, which is expressed as follows:

\begin{equation}
    \text{MPT}(\mathbf{X})=\mathbf{X}_o+\text{GELU}(\mathbf{X}_o^{\top}\mathbf{W}_{o1})^{\top}\mathbf{W}_{o2}+\mathbf{b}_{o2}\text{,}
    \label{eq: eq7}
\end{equation}
where,
\begin{equation}
    \mathbf{X}_o=\text{BN}(\mathcal{C}(\mathbf{X}_a+\text{LAMHA}(\mathbf{X}_a), \mathbf{X}_c+\text{GCB}(\mathbf{X}_c)))\text{,}
    \label{eq: eq8a}
\end{equation}
\begin{equation}
    \mathbf{X}_a=\text{BN}(\mathbf{X}^{\top}\mathbf{W}_a+\mathbf{b}_a)\text{,}
    \label{eq: eq8b}
\end{equation}
\begin{equation}
    \mathbf{X}_c=\text{BN}(\mathbf{X}^{\top}\mathbf{W}_c+\mathbf{b}_c)\text{.} 
    \label{eq: eq8c}
\end{equation}

\added{Here, $\mathcal{C}(\cdot)$ is the concatenation operation;} $\mathbf{W}_a \in \mathbb{R}^{c \times d_a}$, $\mathbf{b}_a \in \mathbb{R}^{d_a}$, $\mathbf{W}_c \in \mathbb{R}^{c \times d_c}$, $\mathbf{b}_c \in \mathbb{R}^{d_c}$ are learnable parameters for linear transformations before concatenation.

\subsection{Local aggregation}\label{subsec2.5}

\begin{figure}[h]
\centering
\includegraphics[scale=0.85]{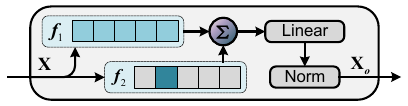}
\caption{Local aggregation module. This module reduces the size in the time dimension, fuses local features through two transformation functions, increases channel depth through a linear projection layer, and uses BatchNorm for normalization.}
\label{fig: fig3}
\end{figure}

\added{In modeling seismic waveforms, we pay closer attention to their temporal variations, recognizing that a single time point cannot adequately capture the rich information contained in seismic data.} Therefore, we proposed a concise and efficient structure called the local aggregation module (Fig. \ref{fig: fig3}) to obtain feature fusion representations within a specified time step in the time dimension.

In this study,  $f_1$ and $f_2$ are implemented using average pooling and max pooling, respectively. \added{Max pooling selectively retains the maximum value among feature points within a local neighborhood, preserving fine-grained details while reducing the overall signal strength of high-frequency components. This approach is particularly effective for segments with abrupt changes in sequences, as it places greater emphasis on low-frequency feature components. }\cite{RN381_0912_37, RN383_0912_38}. However, since max pooling only focuses on the maximum value, it may overlook other important information. \added{In contrast, average pooling prioritizes the preservation of background information and maintains the overall distribution of frequency components. However, due to its equal weighting of all feature points, average pooling may introduce some blurring of detailed information, especially in applications where maintaining fine-grained details is crucial.}

Hence, we proposed the local aggregation Module, which could simultaneously retain background information and enhance detailed features. This module effectively reduces the temporal dimension of features while increasing the feature depth. The computation process for this module can be represented as:

\begin{equation}
    \text{LA}(\mathbf{X},k)=\text{BN}([\mathbf{x}_0,\mathbf{x}_1,\cdots,\mathbf{x}_n]^{\top}\mathbf{W}_a+\mathbf{b}_a)\text{,}
    \label{eq: eq9}
\end{equation}
where $\mathbf{W}_a \in \mathbb{R}^{\mathbf{c}_d \times \mathbf{c}_{d+1}}$, $\mathbf{b}_a \in \mathbb{R}^{c_{d+1}}$; and $c_d$ represents the number of channels of the $d$-th stage; $\mathbf{x}_i$ is a column vector representing the aggregated output within a specified time step. The dimension reduction in the time dimension essentially involves weighted fusion with a strong inductive bias. The weighting method and the calculation of weights $w_j$ for time points $j$ are as follows:
\begin{equation}
    \mathbf{x}_i= \textstyle \sum_{j=i\cdot k}^{i \cdot k + k -1}w_{j}\mathbf{X}_{:,j}\text{,}
    \label{eq: eq10}
\end{equation}
where,
\begin{equation}
w_j=\begin{cases}
        (1+k)/k & j \in M \\
        1/k & \text{otherwise,}
    \end{cases}
    \label{eq: eq11a}
\end{equation}

\begin{equation}
M=\{m_j\mid m_i=\text{argmax}(\mathbf{X}_{:,i\cdot k:i\cdot k + k})\}\text{.}
    \label{eq: eq11b}
\end{equation}

Here, $i=0,1,2,\cdots,\lceil l/k\rceil-1 $.

\section{Data and Results}\label{sec3}
\subsection{Data and labeling}\label{subsec3.1}

\begin{figure*}[thb]
    \centering
    \subfloat []{
        \includegraphics[scale=0.82]{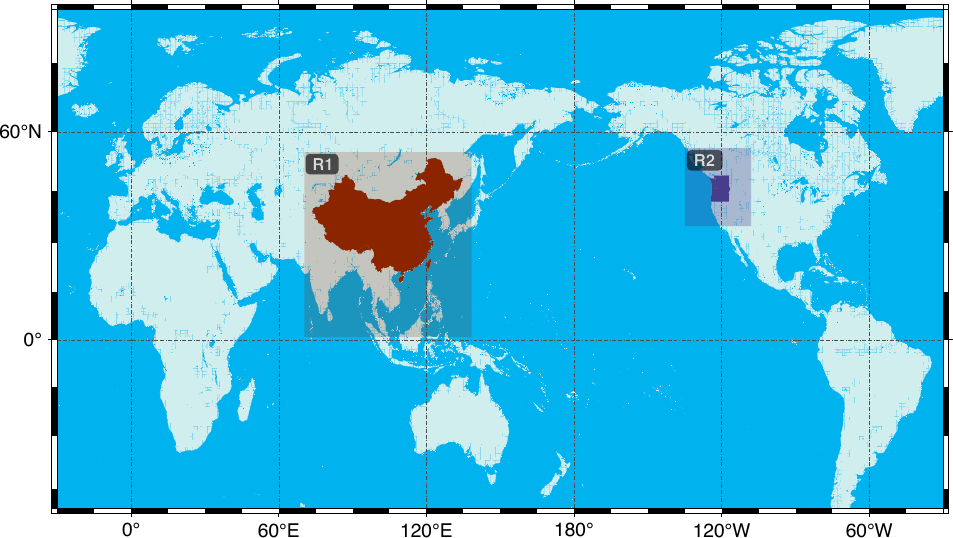}
        }
    \vspace{0.1cm}
    \subfloat []{
    \hspace{0.3cm}
        \includegraphics[scale=0.82]{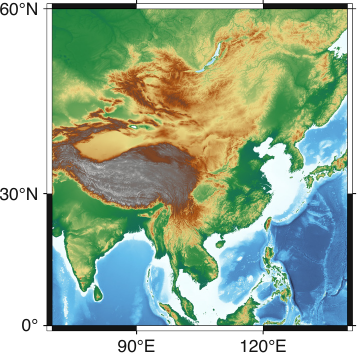}
        }
    \hspace{1cm}
    \subfloat []{
        \includegraphics[scale=0.82]{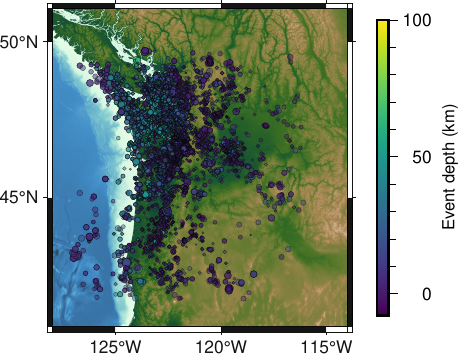}
        }
    \caption{Geographical distribution of DiTing and PNW datasets. (a) Comparison of seismic event distribution in both datasets. (b) Distribution of seismic events in the DiTing dataset (R1). (c) Visualization of seismic event distribution (R2), magnitude, and source depth in the PNW dataset.}
    \label{fig: fig4}
\end{figure*}

We utilized the DiTing dataset \cite{RN261_0912_20} from the China region and the PNW dataset \cite{RN20_0912_21} from the Northwestern Pacific region of the United States. Both datasets comprise a substantial number of seismic events, including arrival times, first-motion polarity, and magnitude, among other labels. Notably, \added{there are no overlapping seismic events in the two datasets}. The geographical distribution of the two datasets is illustrated in Fig. \ref{fig: fig4}. The distributions of signal-to-noise ratio (SNR), epicentral distance, and magnitude are shown in Supporting Figure S3.

We employed the DiTing dataset for training our neural network model. \added{DiTing is a large-scale dataset obtained from the China Seismic Network (CSN), collected and organized by the Institute of Geophysics, China Earthquake Administration. It encompasses seismic events in mainland China and its neighboring regions spanning from 2013 to 2020.} The magnitudes of these seismic events range from 0 to 7.7, with epicentral distances ranging from 0 to 330 kilometers. The waveform data have a length of 180 seconds and a sampling rate of 50 Hz. In this study, seismic event samples with complete labels were exclusively utilized to cater to the training requirements for multiple tasks. The dataset was randomly partitioned into a training set (80\%), a validation set (10\%), and a test set (10\%). \added{This encompassed approximately 277k three-component seismic waveform data samples from the China region.}

\added{To assess the cross-regional out-of-distribution generalization ability of the model}, \added{we conducted testing using the PNW dataset from the Pacific Northwest Seismic Network (PNSN).} The "ComCat event" subset of the PNW dataset consists of data verified by analysts and sent to the United States Geological Survey (USGS), encompassing a total of 65,384 events spanning from 2002 to 2022. The seismic waveform data have a length of 150 seconds and a sampling rate of 100 Hz. To maintain consistency with the DiTing dataset labels, we retained samples with clear first-motion polarity and the ML magnitude type. This subset of data was used exclusively as a test set to evaluate the generalization capabilities of the model, thus not included in the training process.

During the training process, the earthquake detection task labels are represented using 0-1 vectors. These vectors start at the P-wave arrival time and end at a specified time denoted as $P_{arrival}+\lambda(S_{arrival}-P_{arrival})$, where in this study, $\lambda=2$. As for the phase-picking task, a probability sequence is employed to indicate the probability of phase arrival times. We tested four different label forms: rectangular, triangular, Gaussian, and Sigmoid, to smooth the phase arrival positions \cite{RN365_0912_22}. In our hyperparameter selection process, the Gaussian label yielded lower loss and higher F1-Score, justifying its being applied in the final model. \added{Under this label format, the probability of P-wave and S-wave arrival times was set to 1 at the manually marked positions and gradually decreased to 0 before and after that sampling point to follow a Gaussian distribution, with a total width of 0.5 seconds. For the first-motion polarity classification task, we utilized one-hot vectors to represent the first-motion polarity as upward and downward.}

\subsection{Evaluation}\label{subsec3.2}

Various metrics were employed to evaluate the performance of the models on different tasks. \added{The evaluation metrics chosen include Precision (Pr), Recall (Re), F1-Score (F1), Mean Error (Mean), Standard Deviation (Std.), Mean Absolute Error (MAE), and coefficient of determination (R\textsuperscript{2}),} defined as follows:

\begin{equation}
    \text{Pr}=\frac{T_\text{p}}{F_\text{p}+T_\text{p}}\text{,}
    \label{eq: eq12}
\end{equation}

\begin{equation}
    \text{Re}=\frac{T_\text{p}}{F_\text{n}+T_\text{p}}\text{,}
    \label{eq: eq13}
\end{equation}

\begin{equation}
    \text{F1}=\frac{2\times \text{Pr} \times \text{Re}}{\text{Pr} + \text{Re}}\text{,}
    \label{eq: eq14}
\end{equation}

\begin{equation}
    \text{Mean}=\frac{1}{N}\textstyle \sum_{i=1}^{N}(y_i - \hat{y}_i)\text{,}
    \label{eq: eq15}
\end{equation}

\begin{equation}
    \text{Std}=\sqrt{\frac{1}{N}\textstyle \sum_{i=1}^{N}(y_i - \hat{y}_i)^2}\text{,} 
    \label{eq: eq16}
\end{equation}

\begin{equation}
    \text{MAE}=\frac{1}{N}\textstyle \sum_{i=1}^{N}|y_i - \hat{y}_i| \text{,}
    \label{eq: eq17}
\end{equation}

\begin{equation}
    \text{R\textsuperscript{2}}=1-\frac{\textstyle \sum_{i=1}^{N}(y_i - \hat{y}_i)^2}{\textstyle \sum_{i=1}^{N}(y_i - \overline{y_i})^2}\text{,}
    \label{eq: eq18}
\end{equation}
where $T_\text{p}$ represents the number of true positives; $F_\text{p}$ represents the number of false positives; $F_\text{n}$ denotes the number of false negatives; $N$ denotes the total number of samples; $y_i$ represents the labels of sample $i$; $\hat{y_i}$ represents the predictions of sample $i$; and $\overline{y}$ represents the mean value of the sample labels. Precision, recall, and F1-Score are common evaluation metrics in classification tasks. In the context of phase-picking tasks, samples with larger residuals are considered false positives. Accordingly, samples with residuals within the error tolerance $\delta<0.1\text{s}$ were chosen as true positives in this study. The F1-Score is a balanced metric between precision and recall, providing a comprehensive evaluation of the 
performance of the model. MAE effectively measures the degree of offset between model predictions and sample true labels. Standard deviation indicates the stability of the predictive performance of the model. MAE is used to represent the magnitude of errors between model predictions and labels. The coefficient of determination (R\textsuperscript{2}) is used to measure the correlation between predicted values and true values in regression tasks.

\subsection{Training details}\label{subsec3.3}

When training SeisT, we initialized the linear and convolutional layers using truncated normal distribution \cite{RN385_0925_39}, set BatchNorm weights to 1, and initialized all biases to 0. We utilized the cyclic learning rate scheduler \cite{RN268_0912_24} with the Adam optimizer \cite{RN387_0912_40}, setting the minimum learning rate to $8\times 10^{-5}$ and the maximum learning rate to $1\times 10^{-3}$. Periodic learning rate adjustments help prevent the model from getting stuck in local minima or saddle points. Early stopping was configured as follows: if the loss on the validation set did not decrease for 30 consecutive epochs, training was halted.

We employed data augmentation strategies on the training dataset, including adding random Gaussian white noise, time drift, introducing gaps, channel dropout, amplitude scaling, pre-emphasis, and generating noise with probabilities of 0.4, 0.4, 0.4, 0.4, 0.4, 0.97, and 0.05, respectively, in seismic waveforms. Regularization methods such as DropPath \cite{RN389_0912_41} and Dropout \cite{RN391_0912_42} were used with probabilities of 0.1, 0.2, and 0.3 for models S, M, and L, respectively.

We trained the models for seismic phase picking, earthquake detection, first-motion polarity classification, magnitude estimation, back-azimuth estimation, \added{and epicentral distance estimation} using the PyTorch framework on a single NVIDIA RTX-4090 GPU. \added{The training time varies from 4 to 18 hours depending on model size and application task}. To ensure a fair performance comparison among different models, the same data augmentation methods, optimizers, and learning rates were used for training models on the same tasks.

\subsection{Results}\label{subsec3.4}
\subsubsection{Comparison with other methods}\label{subsubsec3.4.1}

\begin{figure*}[!ht]

    \centering
    \subfloat []{
        \includegraphics[width=7.2cm]{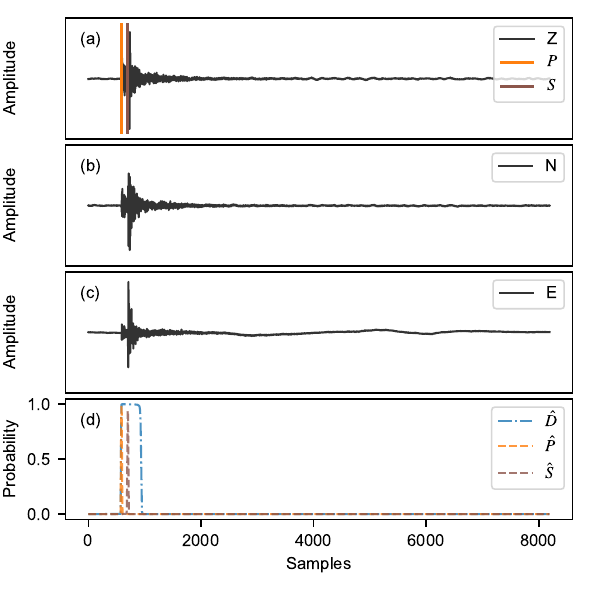}
        }
    \hspace{0.1cm}
    \subfloat []{
        \includegraphics[width=7.2cm]{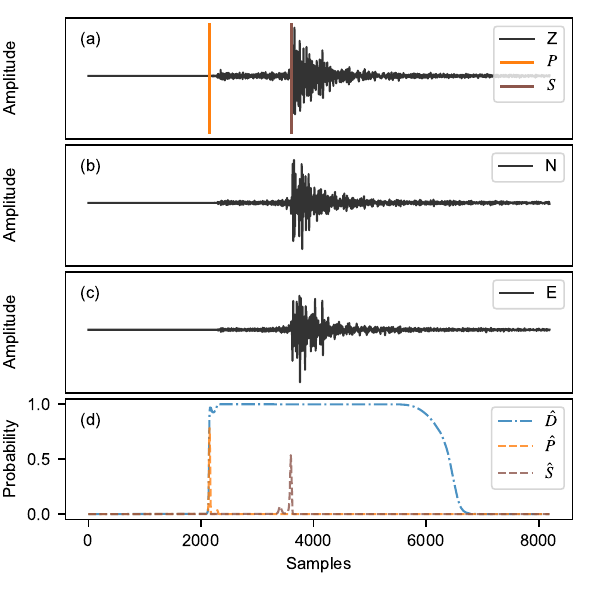}
        }
    \vspace{0.1cm}
    \subfloat []{
        \includegraphics[width=7.2cm]{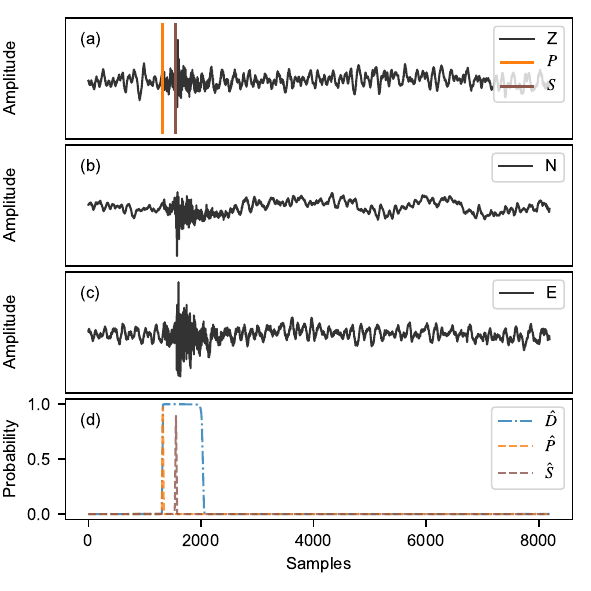}
        }
    \hspace{0.1cm}
    \subfloat []{
        \includegraphics[width=7.2cm]{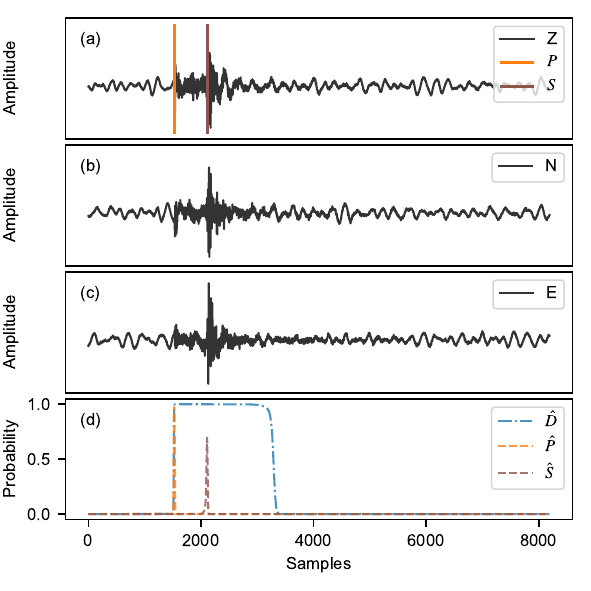}
        }
    \caption{Phase picking and earthquake detection examples. (a-d) are event examples from the test dataset, encompassing various SNRs, epicentral distances, and magnitudes. \added{Specifically, (a) and (b) exhibit relatively high SNR values of 29 dB and 43 dB, respectively. The corresponding event of (a) has a magnitude of 1.7 and an epicentral distance of 37.5 km, and the corresponding event of (b) has a magnitude of 2.1 and an epicentral distance of 98 km. (c) and (d) exhibit lower SNR values of 4 dB and 1 dB, respectively. The corresponding event of (c) has a magnitude of 1.5 and an epicentral distance of 16 km, and the corresponding event of (d) has a magnitude of 5.0 with an epicentral distance of 216 km. Within the context of these examples, Z, N, and E represent the seismogram components. $P$ and $S$ represent the manual picking labels of P-wave and S-wave arrivals, respectively. $\hat{P}$ and $\hat{S}$ represent the predicted probabilities for P-wave and S-wave arrivals, respectively. $\hat{D}$ represents the probability of predicting the time from the P-wave arrival to the end of the S-wave coda.}}
    \label{fig: fig5}
\end{figure*}

Seismograms of the “ENZ” (East, North, Vertical) components.

\begin{table*}[ht]
\caption{Model performance comparison}
\label{table: model_performance_comparison}
\centering
\begin{tabular}{llccccccc}
\toprule
Task & Model & F1 & Pr & Re & Mean & Std. & MAE & R\textsuperscript{2}  \\ 
\midrule

\multirow{5}{*}{Phase-P} 
 & PhaseNet      & 90.23 & \textbf{96.91} & 84.41 & 0.02 & \textbf{0.03} & \textbf{0.02} & - \\
 & EQTransformer & 93.96 & 95.59 & 92.38 & 0.02 & \textbf{0.03} & 0.03 & - \\ 
 & SeisT-S       & 95.66 & 96.70 & 94.64 & 0.02 & \textbf{0.03} & \textbf{0.02} & - \\
 & SeisT-M       & 95.63 & 96.85 & 94.44 & 0.02 & \textbf{0.03} & \textbf{0.02} & - \\
 & SeisT-L       & \textbf{95.73} & 96.62 & \textbf{94.85} & \textbf{0.01} & \textbf{0.03} & \textbf{0.02} & - \\ 
\hline
\multirow{5}{*}{Phase-S} 
 & PhaseNet      & 57.54 & 69.06 & 49.32 & \textbf{0.02} & \textbf{0.04} & \textbf{0.02} & - \\
 & EQTransformer & 58.40 & 62.10 & 55.12 & 0.03 & \textbf{0.04} & 0.03 & - \\ 
 & SeisT-S       & 64.16 & 67.55 & 61.09 & 0.03 & 0.05 & 0.03 & - \\
 & SeisT-M       & 67.42 & 70.04 & 64.98 & 0.03 & 0.05 & 0.03 & - \\
 & SeisT-L       & \textbf{67.98} & \textbf{70.83} & \textbf{65.35} & 0.03 & 0.05 & 0.03 & - \\ 
\hline
\multirow{4}{*}{Detection} 
 & EQTransformer & 95.57 & 96.04 & \textbf{95.10} & - & - & -  & - \\
 & SeisT-S       & 95.13 & \textbf{97.38} & 92.99 & - & - & - & - \\ 
 & SeisT-M       & 95.43 & 96.43 & 94.45 & - & - & - & - \\
 & SeisT-L       & \textbf{95.60} & 96.71 & 94.52 & - & - & - & - \\
 \hline
\multirow{4}{*}{First-Motion} 
 & DitingMotion & 94.26 & 94.40 & 94.14 & - & - & - & -  \\
 & SeisT-S      & 93.96 & 94.11 & 93.85 & - & - & - & - \\
 & SeisT-M      & \textbf{94.70} & \textbf{94.79} & \textbf{94.64} & - & - & - & - \\
 & SeisT-L      & 94.36 & 94.39 & 94.34 & - & - & - & - \\
\hline
\multirow{4}{*}{Magnitude} 
& MagNet   & - & - & - & \added{\textbf{0.00}}   & \added{0.30}  & \added{0.22}  & \added{0.79}\\
& SeisT-S  & - & - & - & \textbf{0.00}   & 0.24  & 0.18  & 0.86 \\
& SeisT-M  & - & - & - & \textbf{0.00}   & \textbf{0.23}  & \textbf{0.17}  & \textbf{0.87} \\
& SeisT-L  & - & - & - & 0.01   & 0.24  & 0.18  & 0.86 \\
\hline
\multirow{4}{*}{Back-Azimuth} 
 & Baz-Network  & - & - & - & \textbf{0.07}   & \textbf{69.47}  & \textbf{43.89}  & \textbf{0.55} \\
 & SeisT-S      & - & - & - & -0.23  & 70.18  & 45.09  & 0.54 \\
 & SeisT-M      & - & - & - & 0.40   & 69.85  & 43.95  & 0.54 \\
 & SeisT-L      & - & - & - & -0.14  & 69.56  & 43.96  & \textbf{0.55} \\
 \hline
\multirow{3}{*}{\added{Distance}} 
 & \added{SeisT-S}      & \added{-} & \added{-} & \added{-} & \added{-4.37}  & \added{13.29}  & \added{5.27}  & \added{0.78} \\
 & \added{SeisT-M}      & \added{-} & \added{-} & \added{-} & \added{\textbf{-3.81}}  & \added{\textbf{12.23}}  & \added{\textbf{4.80}}  & \added{\textbf{0.81}} \\
 & \added{SeisT-L}      & \added{-} & \added{-} & \added{-} & \added{-4.01}  & \added{12.72}  & \added{5.03}  & \added{0.80} \\
\bottomrule
\multicolumn{9}{p{14cm}}{F1, P, and R are F1-score, precision, and recall, respectively. Mean and Std are the mean and standard deviation of errors in seconds respectively. MAE is the mean absolute error. R\textsuperscript{2} is the coefficient of determination. The training and testing of the magnitude estimation task were performed on the PNW dataset. The training and testing of the back-azimuth estimation task were performed on the DiTing dataset. \added{The phase-picking, earthquake detection, first-motion polarity classification, and epicentral distance estimation} tasks are trained using the DiTing dataset, and tested on the PNW dataset. \added{Bold values represent the best performance.}}
\end{tabular}
\end{table*}

To comprehensively evaluate model performance, we conducted experiments across multiple tasks, \added{including phase picking, earthquake detection, first-motion polarity classification, magnitude estimation, back-azimuth estimation, and epicentral distance estimation}. To be exact, the top-performing models currently available were selected, including PhaseNet \cite{RN266_0912_7}, EQTransformer \cite{RN11_0912_5}, DiTingMotion \cite{RN255_0912_6}, MagNet \cite{RN17_0912_12}, and Baz-Network \cite{RN51_0912_14}. To ensure fairness and comparability of experiments, all these models were trained on the same training dataset and evaluated on the same test dataset. For each task, different models were trained using consistent training strategies and hyperparameters. It is worth emphasizing that we did not apply additional filtering to the test data to maintain the originality of the data. During the model training process, we employed various data augmentation techniques such as random cropping, time shifting, and amplitude scaling \cite{RN367_0912_23}. \added{These techniques contributed to improving the convergence speed of the model, reducing the risk of overfitting, and improving the generalization of the trained model. We applied the same data augmentation strategies for both our study and the baseline models to ensure a fair comparison. Before inputting the data into the model, we normalized the waveforms to scale the amplitudes within the range suitable for the processing of the model.}

In multiple tasks, the SeisT model has demonstrated superior performance, surpassing existing models or achieving comparable performance with them. The phase-P and phase-S picking tasks can be taken as examples. We compared SeisT with state-of-the-art deep learning models such as EQTransformer and PhaseNet (see Table \ref{table: model_performance_comparison}). Six metrics, namely, Precision, Recall, F1-Score, Mean, Std, and MAE, were employed to evaluate model performance, where predictions with absolute errors smaller than the error tolerance were considered as true positives. An in-depth discussion and analysis of the relationship between model performance and error tolerance is also provided in Section \ref{sec4}. Due to significant feature differences between the PNW test dataset and the DiTing dataset, they do not adhere to the assumption of independent and identically distributed (i.i.d.) data, making the PNW test dataset highly effective in assessing the out-of-distribution generalization capability of the model. The test results on the i.i.d. dataset are shown in Supporting Table S3. On the PNW test dataset, compared with other models, the SeisT model demonstrates significant performance improvements. For the phase-P picking task, all three SeisT variants exhibit similar performance and achieve F1-Score improvements of 5.5\% and 1.7\% over PhaseNet and EQTransformer, respectively. Notably, the lightweight SeisT-S model utilizes only half of the PhaseNet parameters and one-third of the EQTransformer parameters. For the phase-S picking task, SeisT-M and SeisT-L similarly outperform existing models, achieving over 9\% performance improvement over existing models. Even the lightweight SeisT-S achieves a performance improvement of over 5\%. As is observed, the phase-P picking task typically exhibits higher precision than the phase-S picking task. \added{This is because the P-waves coda partially overlaps with the initial arrivals of S-waves}, making it more complex and challenging to accurately capture the features of S-waves. This in turn makes the phase-S picking task more challenging. In earthquake detection tasks, SeisT performs at a similar level to existing models across various evaluation metrics. \added{Some examples of time picking and earthquake detection are shown in Fig. \ref{fig: fig5}.} In the first-motion polarity classification task, SeisT achieves performance metrics similar to DiTingMotion, indicating their comparable performance levels. In the magnitude estimation task, metrics such as Mean, Std, MAE, and R\textsuperscript{2} were used to evaluate model performance, and the results show that SeisT outperforms MagNet 2\% in terms of determination coefficient on the test dataset. However, we also noted that the out-of-distribution generalization capability of the model is not outstanding in the magnitude estimation task. \added{This may be due to the differences in seismic networks, geological features, magnitude types, and magnitude sizes in different regions.} In the back-azimuth estimation task, the existing Baz-Network model requires the computation of covariance matrices, eigenvalues, and eigenvectors, followed by feature fusion through convolution and fully connected layers. This results in a parameter count exceeding 1050k, significantly higher than those of the SeisT series, namely, 98k, 312k, and 529k. Unexpectedly, SeisT achieves comparable or even superior performance with fewer parameters in the testing phase compared to Baz-Network. \added{In the epicentral distance estimation task, all three models of SeisT achieved the R\textsuperscript{2} of 0.98 on the DiTing test set. In cross-regional out-of-distribution generalization performance testing, the R\textsuperscript{2} of three models reached 0.78, 0.81, and 0.80, respectively, and the MAE was only 5.27, 4.80, and 5.03, respectively.}

\subsubsection{Ablation study}\label{subsubsec3.4.2}

We conducted ablation experiments to understand the contributions of the main modules in SeisT, as well as the effects of parameter count and training schedules. These experiments were conducted on the DiTing dataset, focusing on time picking and earthquake detection tasks. \added{The following studies were performed: (a) Ablation study of MSMC module. (b) Ablation study of LA module. (c) Ablation study of MPT module.} (d) Performance of SeisT variants with different parameter counts. (e) Performance of SeisT under different training schedules.

According to the ablation experimental results (Supporting Table S2), the MSMC module demonstrates a noticeable superiority over the regular convolution module, especially in the phase-P picking task, where its F1-Score improves by 1\%, compared with the regular convolution module. The MSMC module, based on convolutional kernels of multiple sizes, is applied across different channels. Its superiority lies in its ability to extract rich multi-level features across multiple channels, effectively capturing underlying feature information and reducing the influence of irrelevant features on the results. \added{Furthermore, The LA module is a simple yet effective structure. Experiments demonstrate that when employing a down-sampling approach without the LA module, its performance decreases by 0.8\% to 2.7\% in earthquake detection and seismic phase picking tasks.}

\added{Compared to conventional self-attention mechanisms, the MPT module effectively mitigates computational burden in earthquake detection tasks while demonstrating notable performance enhancements.} When considering the model without the attention mechanism path, the F1-Score for earthquake detection is only 88.98\%. However, introducing the MPT module into the SeisT model leads to nearly an 8\% performance improvement. This performance boost is attributed to the capability of the global self-attention mechanism to effectively capture long-range dependencies, which traditional convolution or recurrent neural networks struggle to capture. Furthermore, the attention mechanism brings about a 3\% and 5\% performance improvement for phase-P and phase-S picking tasks, respectively.

We tested models of three different sizes, S, M, and L, with the main differences being the feature map sizes and the number of layers within each stage. As expected, SeisT-L achieved the best performance on the DiTing dataset, while SeisT-S exhibited slightly lower performance metrics compared with SeisT-L, and SeisT-M fell in between the two. In the phase-P picking task, the F1-Score of SeisT-M is 0.02\% higher than that of SeisT-L. Although this small performance difference may be negligible, it could be attributed to SeisT-L having a larger number of parameters, potentially causing it to slightly overfit the training data and thus achieving performance close to SeisT-M, which has only half the parameter count.

In a thorough exploration of hyperparameter choices, we observed that cyclic learning rate schedules \cite{RN268_0912_24}, as opposed to linear decay or fixed learning rates, could significantly improve model performance. \added{Cyclic learning rate schedules may help prevent the model from becoming trapped in local minima or saddle points, potentially leading to improved performance.} As highlighted by Dauphin et al. \cite{RN369_0912_25}, compared with local minima, saddle points hinder convergence more, and higher learning rates help to quickly escape saddle points. This is precisely the effect achieved by periodically increasing the learning rate.

\section{Discussion}\label{sec4}
\added{The SeisT model proposed in this paper demonstrates competitive or superior performance compared to existing state-of-the-art models in a variety of seismic monitoring tasks, including earthquake detection, seismic phase picking, first-motion polarity classification, magnitude estimation, back-azimuth estimation, \added{and epicentral distance estimation}. Its strong performance is primarily attributed to the model architecture design and data augmentation methods used during training.} 

The design of the MSMC and MPT plays a crucial role in extracting potential information from the \added{seismograms}. The self-attention mechanism with local aggregation empowers the neural network to focus on both local and global features simultaneously. \added{The availability of the DiTing dataset provides ample training data, and the use of the PNW dataset allows for assessing cross-regional out-of-distribution generalization performance, particularly focusing on different regional distributions.}

\begin{figure}[h]

    \centering
    \subfloat []{
        \includegraphics[width=7cm]{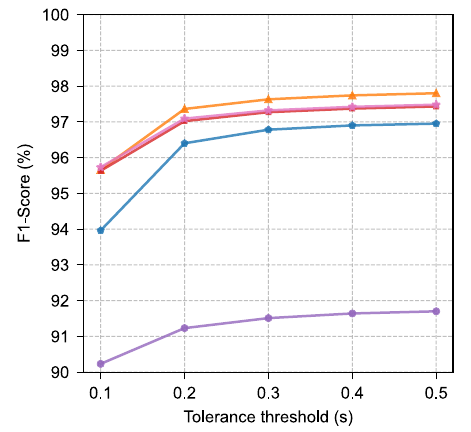}
        }
    \vspace{0.1cm}
    \subfloat []{
        \includegraphics[width=7cm]{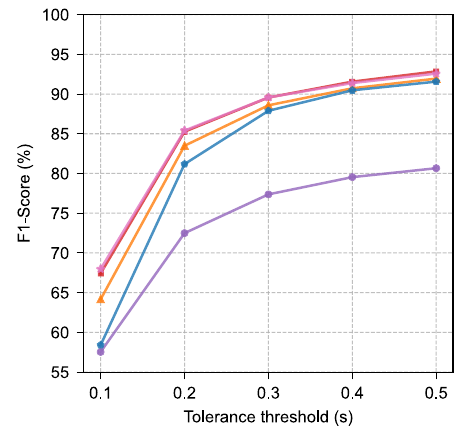}
        }
    \vspace{0.1cm}
    \subfloat{
        \includegraphics[width=6cm]{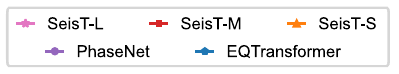}
        }
    \caption{Performance comparison of seismic phase picking among different models. \added{(a) corresponds to the performance of phase-P picking, while (b) corresponds to the performance of phase-S picking. Samples with an absolute error less than the tolerance threshold (from 0.1s to 0.5s) are considered true positives. The models were trained on the DiTing dataset and evaluated on the PNW dataset.}}
    \label{fig: fig6}
\end{figure}

Within the SeisT model, the stem module is designed for preprocessing input seismograms. This module utilizes multi-path depth-wise separable convolutions and linear transformations to rapidly reduce the time dimension of seismograms. This significantly reduces the computational overhead for subsequent convolutional operations and attention mechanisms. Directly inputting the waveforms into multi-head self-attention layers would incur an unacceptable computation complexity of $\mathcal{O}(L^2)$. In the body module, the number of stages can be flexibly adjusted based on application requirements. In general, deeper networks tend to possess stronger non-linear fitting and feature representation capabilities, potentially leading to higher accuracy. However, deeper networks also come with increased computational complexity and a higher risk of overfitting. Therefore, the proposed models adopt a design with 4 stages, a structure utilized by various network models in the field \cite{RN177_0912_26, RN293_0912_27, RN316_0912_28, RN371_0912_29}. This design effectively reduces dimensionality while extracting deep features.

Among numerous tasks, phase picking stands out as one of the most representative tasks. This is due to its requirement to provide probability values for each sample point in the corresponding waveform, which relies on the accurate extraction of seismic wave features. PhaseNet and EQTransformer are widely used phase-picking models. In our out-of-distribution generalization tests, we observed varying degrees of performance degradation in phase-P and phase-S picking for both models when using low error tolerance. Notably, S-wave performance witnessed a significant decline, with a decrease of over 20\% in maximum F1-Score when the tolerance threshold was reduced from 0.2s to 0.1s. This decline can be attributed to the overlap of S-wave with P-wave arrivals, making accurate phase-S picking at low tolerance thresholds exceptionally challenging. Moreover, manual phase picking introduces a certain degree of subjectivity, potentially leading to poor model performance at low error tolerance settings.

\begin{figure*}[!htb]

    \centering
    \subfloat []{
        \includegraphics[width=6cm]{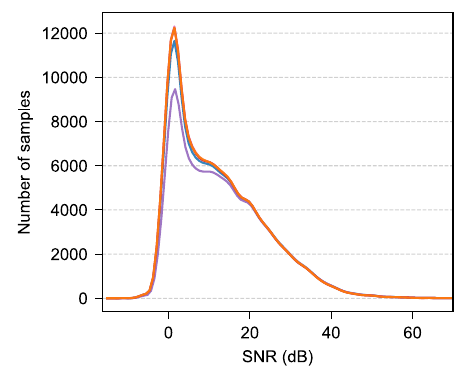}
        }
    \hspace{0.5cm}
    \subfloat []{
        \includegraphics[width=6cm]{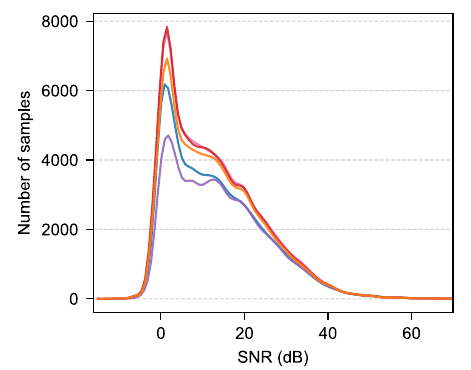}
        }
    \vspace{0.01cm}
    \subfloat []{
        \includegraphics[width=6cm]{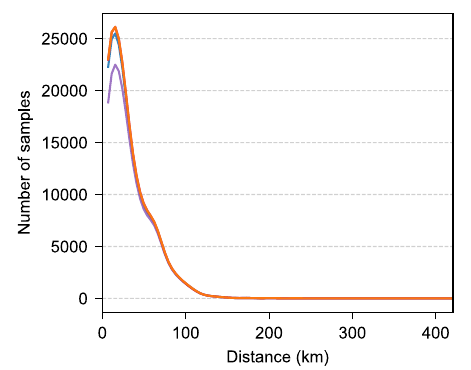}
        }
    \hspace{0.5cm}
    \subfloat []{
        \includegraphics[width=6cm]{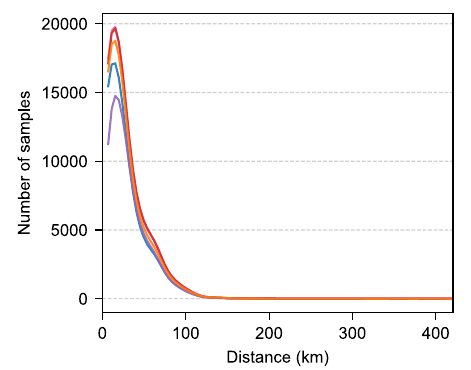}
        }
    \vspace{0.01cm}
    \subfloat []{
        \includegraphics[width=6cm]{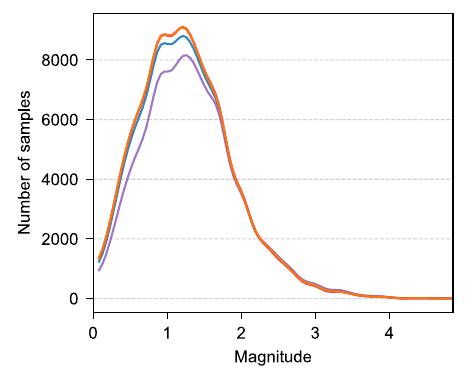}
        }
    \hspace{0.5cm}
    \subfloat []{
        \includegraphics[width=6cm]{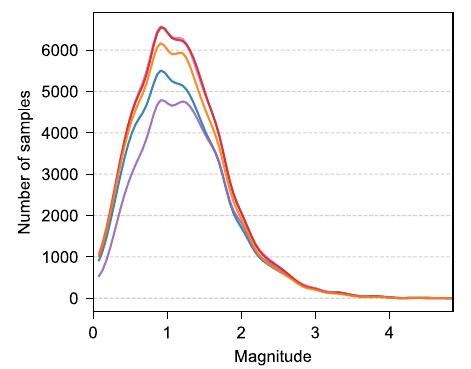}
        }
    \vspace{0.01cm}
    \subfloat{
        \includegraphics[width=10cm]{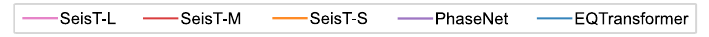}
        }
    \caption{Comparison of phase picking performance for earthquake events with different SNRs, epicentral distances, and magnitudes.  (a) and (b) represent the performance distribution of various phase-picking models under different SNR conditions for phase-P picking and phase-S picking, respectively. (c) and (d) represent the performance distribution of various phase-picking models under different epicentral distance conditions for phase-P picking and phase-S picking, respectively. (e) and (f) represent the performance distribution of various phase-picking models under different magnitude levels for phase-P picking and phase-S picking, respectively. The test data utilized in this analysis are derived from the PNW dataset.}
    \label{fig: fig7}
\end{figure*}

Even under the demanding error tolerance threshold of 0.1s, compared with other picking models, SeisT exhibited significantly better performance. Specifically, SeisT-M outperformed EQTransformer by 1.7\% and 9.0\% in P-wave and S-wave arrival time picking, respectively, on the out-of-distribution test set. SeisT consistently showed superior performance across different tolerance thresholds (ranging from 0.1s to 0.5s) on the out-of-distribution test set, as demonstrated in Fig. \ref{fig: fig6}. This suggests that the SeisT model adapts well to \added{seismic waveforms} monitored by different networks in various regions, exhibiting robust out-of-distribution generalization capabilities. Such capabilities significantly reduce the transfer learning cost when the model is to be applied in practice.

\begin{figure*}[!htb]

    \centering
    \subfloat []{
        \includegraphics[width=16cm]{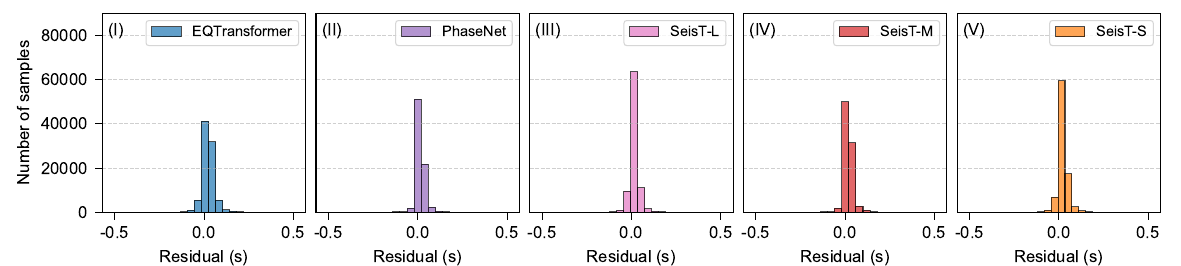}
        }
    \vspace{0.1cm}
    \subfloat []{
        \includegraphics[width=16cm]{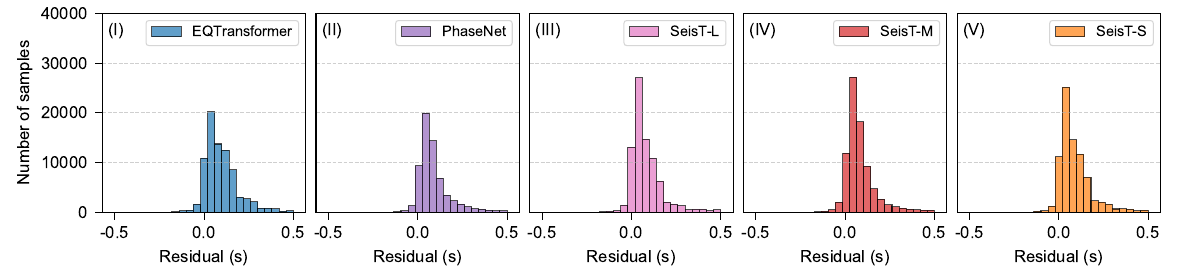}
        }
 
    \caption{Comparison of residual distributions between multiple models for automated phase picking and manual picking. (a) and (b) represent the residual distribution of P-wave and S-wave arrival time picking, respectively. The test data used are from the PNW dataset.}
    \label{fig: fig8}
\end{figure*}

\added{The ability to pick arrivals is correlated with the SNR, epicentral distance, and magnitude of seismic waveforms.} As shown in Fig. \ref{fig: fig7}, the relationship between the number of P-wave and S-wave picks in different models and the SNR, epicentral distance, and magnitude is depicted. While this distribution trend is determined by the distribution of the data samples themselves, the disparities observed across different models can still reflect variations in model performance. The application of phase-picking models on low signal-to-noise ratio waveforms has long been a challenge. Compared with the baseline model, the SeisT series models exhibit noticeable performance improvements in low signal-to-noise ratio conditions, especially in the signal-to-noise ratio range of 0-10 dB, demonstrating significant differences. Under different epicentral distances, substantial differences are mainly manifest in the phase-S picking of seismic events within the 0-50 km range. Different seismic magnitudes can exhibit variations in signal characteristics, with smaller magnitude events implying weaker signal strength, potentially increasing the difficulty of signal analysis. In the low-magnitude range, SeisT continues to outperform other models, indicating its higher sensitivity and ability to ensure picking performance even for weaker signals. Fig. \ref{fig: fig8} illustrates the residual distribution of different models in phase-P and phase-S picking tasks. In the vicinity of zero, the SeisT series performs the best, with PhaseNet slightly outperforming EQTransformer in phase-P picking and demonstrating similar performance in phase-S picking. One possible reason for this is that extensive residual connections of PhaseNet allow it to retain a substantial amount of original fine-grained features, thus achieving better performance in the relatively straightforward phase-P picking task.

\begin{figure}[!htb]

    \centering
    \subfloat []{
        \includegraphics[width=6.7cm]{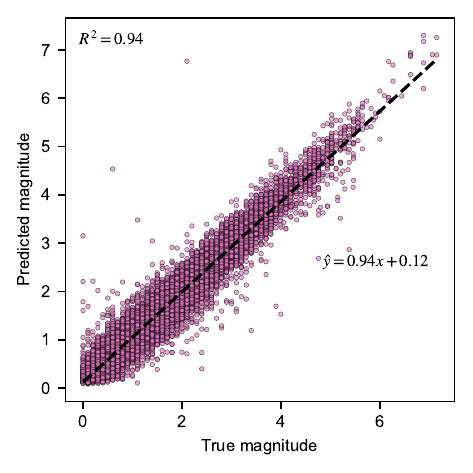}
        }
    \vspace{0.1cm}
    \subfloat []{
        \includegraphics[width=6.7cm]{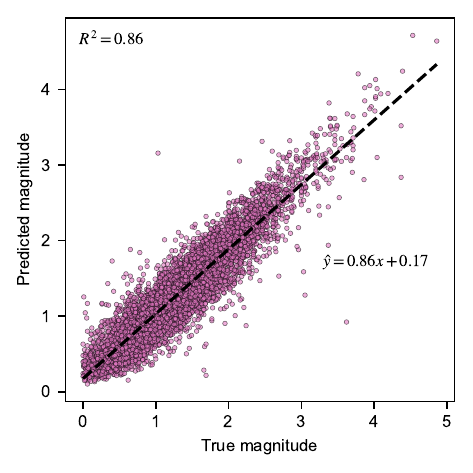}
        }

    \caption{The performance of SeisT in the magnitude estimation task. (a) represents the performance of SeisT-L on the DiTing test dataset, achieving R\textsuperscript{2} of 0.94. (b) denotes the performance of SeisT-L on the PNW test dataset, with the R\textsuperscript{2} value of 0.86.}
    \label{fig: fig9}
\end{figure}

\begin{figure}[!ht]
\centering
\includegraphics[width=8cm]{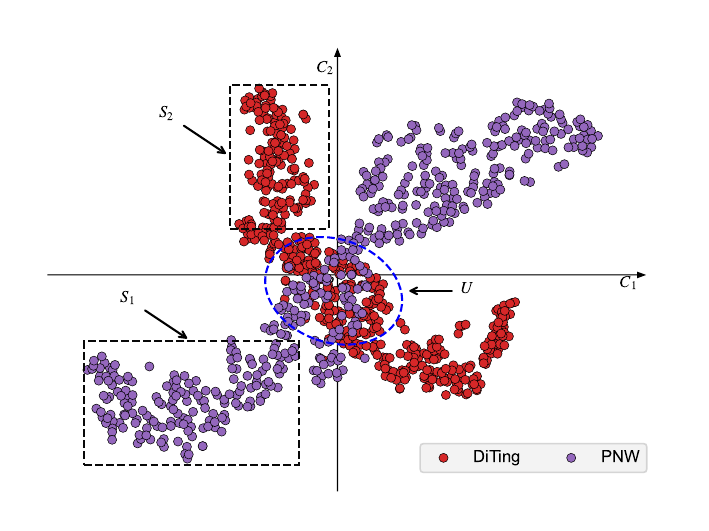}
\caption{t-SNE visualization of data sample feature representation. Randomly extracting an equal number of samples from the DiTing and PNW datasets, we visualized their feature representations using t-SNE. Here, $U$ denotes that a small subset of samples from both datasets exhibits similarity in the high-dimensional complex feature space. $S_1$ and $S_2$ indicate that the majority of samples from the two datasets are distinctly separable in the high-dimensional complex feature space, signifying significant feature differences. $C_1$ and $C_2$ represent two feature dimensions.}
\label{fig: fig10}
\end{figure}

First-motion polarity classification is a classification task that primarily relies on waveform characteristics in the immediate aftermath of the P-wave arrival. Therefore, the model needs to input a segment of the waveform containing the P-wave arrival time and a period afterward. First-motion polarity is closely related to the trend of waveform changes following the P-wave arrival. DiTingMotion introduces an additional input channel, which is the amplitude difference between adjacent time points. This method can to some extent reduce the learning difficulty of waveform trends for the neural network. SeisT, on the other hand, does not adopt this approach. \added{To maintain simplicity and computational efficiency, SeisT utilizes the normalized three-component seismic waveforms as input for the first-motion polarity classification task, without the need for any additional pre-computed data. Despite its straightforward approach, SeisT achieves comparable performance to DiTingMotion.}

In the seismic magnitude estimation task, the data distribution unexpectedly influences the models. Unlike several other tasks, seismic magnitude estimation requires training the model on data with similar distributions. For example, it is difficult to apply a model trained on data collected from the CSN to seismic waveforms collected from the PNSN due to differences in seismic network equipment. In this study, we separately trained SeisT on the DiTing and PNW datasets and evaluated them on their respective test sets. \added{On the PNW test set, the R\textsuperscript{2} values for SeisT-S, SeisT-M, and SeisT-L are 0.86, 0.87, and 0.86, respectively. Fig. \ref{fig: fig9} displays the relationship between the predicted and true magnitudes of SeisT-L on the DiTing and PNW test sets. MagNet performs similarly to SeisT on the DiTing dataset, with a coefficient of determination of 0.91 (see Supporting Table S3), and on the PNW dataset, it achieves the R\textsuperscript{2} of 0.79, which is 0.07 to 0.08 lower than the SeisT series.}

\added{Due to the absence of official back-azimuth labels in the PNW dataset, the training and testing for the back-azimuth estimation task were solely performed using the DiTing dataset.} In current applications, back-azimuth estimation often occurs after seismic source localization. Therefore, back-azimuth estimation based on deep neural networks helps rapidly determine back-azimuth angles, providing a valuable reference for subsequent work. Testing on the DiTing dataset reveals that Baz-Network and SeisT perform similarly in back-azimuth estimation, with coefficients of determination of approximately 0.54 and MAEs exceeding 40. While these results may not provide high-precision back-azimuth estimates, they still keep the average error within a sharp angle range. Using this model as a preliminary task for localization can contribute to the reliability of earthquake localization results.

\added{The epicentral distance estimation is a typical regression task. In previous studies, it is often closely related to travel times and seismic source depth. In this study, we only utilize seismograms from a single station and epicentral distance labels for the regression task. The neural network, leveraging its powerful representation learning capability, effectively captures the correlation between seismograms and epicentral distance. In cross-regional out-of-distribution testing, the evaluation metrics R\textsuperscript{2} for the three SeisT models are 0.78, 0.81, and 0.80, respectively. On the test set within the same region as the train set, all three models achieve the R\textsuperscript{2} of 0.98 (see Supporting Table S3). It is worth noting that the task is sensitive to the sampling rate of input seismograms, necessitating the resampling of test data to match the sampling rate of the training data. In the experiments, we found that in the epicentral distance estimation task, when the sampling rates of training and testing data are inconsistent, unacceptable errors in results arise. We attribute this to the model not incorporating travel time information, leading the neural network to be incapable of explicitly learning wave velocities. From a traditional perspective of analyzing travel time differences, when the neural network learns the correlation between epicentral distance and the time differences of P-waves and S-waves, the unknown sampling rate of input seismograms causes the fitting target of the neural network to be the relationship between epicentral distance and the number of sampling points. Consequently, the sampling rate of input seismograms significantly influences the ability of the neural network to perceive the time differences between P-waves and S-waves.}

As previously mentioned, to validate the out-of-distribution generalization ability of the models, we trained each of them on the DiTing dataset, which includes seismic events in China and neighboring regions. Subsequently, \added{we separately tested the models on the DiTing test set and the PNW dataset.} Fig. \ref{fig: fig10} presents the feature distributions of samples from the DiTing and PNW datasets. \added{These features are derived from the output of the SeisT feature extractor, namely the output of the body module.} The t-distributed stochastic neighbor embedding (t-SNE) was applied to reduce the dimensionality of deep seismic waveform features. The $U$ region in the figure indicates that \added{a small portion} of samples from both datasets exhibit some similarity in the high-dimensional complex feature space, making it challenging to distinguish them significantly. The samples in regions $S_1$ and $S_2$ \added{highlight clear differences in the feature distributions for the majority of samples between the two datasets, which may be attributed to significant differences in geological conditions, crustal structures, and source characteristics among different regions. This suggests that the model can effectively differentiate samples from the two datasets using this feature representation, indicating that the model has learned a regional feature representation. Therefore, it can be inferred that there are significant} feature differences between the two datasets, suggesting a lack of independence and identical distribution.

\added{SeisT is designed as a foundational model for seismogram analysis, and its generalization performance on data from different regions has been experimentally validated. With its flexible architecture design, SeisT also holds the potential to address various challenges based on seismograms, such as event classification and seismic source depth estimation. By altering the input channel dimensions, it is also conceivable to apply SeisT to the analysis of seismograms from multiple stations, including earthquake location and focal mechanism inversion. However, our approach still has limitations, primarily manifested in tasks sensitive to sampling rates and amplitudes. The generalization capacity for seismograms with varying sampling rates and amplitudes requires further enhancement. We encourage the exploration of more advanced representation learning methods to endow the model with broader out-of-distribution generalization capabilities.}

\section{Conclusion}\label{sec5}
In this study, we employed a network model with a hybrid design of Transformers and convolution, serving as a foundational model. It can be applied to a variety of tasks, including but not limited to earthquake detection, phase picking, first-motion polarity classification, magnitude estimation, back-azimuth estimation, \added{and epicentral distance estimation} in the fields of seismology. We trained the network model on the DiTing dataset, consisting of seismic data from China and its surrounding regions, using approximately 221k training samples, and evaluated its out-of-distribution generalization performance using 90k test samples of the PNW dataset from the Northwestern Pacific region of the United States. The results indicate that this model exhibits superior out-of-distribution generalization capabilities compared to several existing baseline models. This model holds substantial potential for diverse applications in seismic monitoring, thanks to its robust network architecture, which can play a pivotal role in advancing work related to seismic waveform analysis.

\section*{Code and Data Available}\label{sec6}

Our source code and model are available at \url{https://github.com/senli1073/SeisT} and can reproduce the results presented in this paper. The DiTing \cite{RN261_0912_20} dataset was used for training, validation, and testing and can be obtained at \url{https://www.doi.org/10.12080/nedc.11.ds.2022.0002}. The PNW \cite{RN20_0912_21} dataset was exclusively used for testing the out-of-distribution generalization ability and can be downloaded at \url{https://github.com/niyiyu/PNW-ML}. Maps and figures in this paper were generated using PyGMT \cite{uieda2021pygmt} and Matplotlib \cite{hunter2007matplotlib}. 

\section{Acknowledgements}\label{sec7}
This work was supported in part by the National key
research and development program under Grant 2022YFC3004603, in part by the National Natural
Science Foundation of China under Grant 52274098, and in part by the Natural Science Foundation of Jiangsu Province under Grant BK20221109.

\section{Competing interests}\label{sec9}
The authors declare no competing interests.



\end{document}